# Smart Rewritings of the Basic Equations for Quantitative Non-Linear Inverse Scattering

Martina T. Bevacqua and T. Isernia

*Abstract*—**Non-linearity arising from mutual interactions is one of the two main difficulties to be addressed in inverse scattering. In this paper, we review and describe under a common rationale some approaches which have been introduced in literature in order to counteract non-linearity. In particular, we focus on possible rewritings of the Lippman Schwinger basic equation such to reduce the 'degree of non-linearity' of inverse scattering problem. In detail, three different rewritings are discussed and compared by emphasizing similarities and the differences, and in the same 'rewriting' spirit, we also summarize and discuss the 'Virtual Experiments' framework. Then, some possible joint exploitations of the above concepts are introduced, discussed and tested against numerical examples.**

## I. INTRODUCTION

NON-LINEAR inverse scattering [1]-[3] is a fascinating area for at least two different reasons. First, successful solution approaches to inverse scattering can be of help in very many different applications such as biomedical imaging, subsurface prospecting and non-destructive testing [4]-[6]. Second, because of its non-linearity and ill-posedness, it is very challenging, thus stimulating the efforts and (partial) success of very many researchers and scientists since many years [7]-[11].

In this paper, by reviewing a number of activities performed by a number of researchers in Southern Italy, we focus our attention to issues arising from the non-linearity of the scattering equations with respect to the parameters describing the e.m. characteristics of the target to be retrieved. In fact, it is well known that non-linearity of the problem can induce solution algorithms towards 'false solutions' actually different or even very different from the ground truth [12]. Of course, exploitation of any available a priori information (or even partial information arising from some pre-processing) can considerably reduce the false solution occurrence through convenient starting points or regularization techniques. On the other side, we want to focus herein on the case where no a-priori information is available, so that the solution approach has to rely on pre-processing (if any) and actual quantitative inversion. In fact, we are confident that such an analysis will be indeed anyway of help in all cases when a priori information is available, or global optimization is used (see for instance [13]), or even for the very recent machine learning based approaches (see for instance [14]).

In particular, we focus on nonmagnetic isotropic scatterers. Moreover, as it is simpler to be discussed while still containing (almost) all the relevant elements, we focus herein on the 2D scalar (TM) inverse scattering problem.

The paper takes advantage from two main tools which have been introduced and developed during the last twenty years or so, i.e.:

- the concept of 'degree of non-linearity' of scattering problems with respect to parameters embedding the electromagnetic characteristics of the target [15], and some simple strategies, based on a proper rewriting, such to reduce such a degree [16]-[19];

- the 'Virtual Experiments' framework, where a proper rearranging and assembling of the equations corresponding to the different scattering experiments can allow a conditioning of the internal fields (or contrast sources) which can be then conveniently exploited in the quantitative inversion [20]-[22].

In this paper, we first provide a unitary perspective for the above different contributions. Then, we take advantage from these latter to introduce new effective possibilities. In particular, two different hybrid models and the derived contrast source inversion (CSI) methods [10] are introduced and discussed.

The paper is organized as follows. In Section II, after writing the basic equations in their more standard fashion, we briefly summarize some results regarding the concept of 'degree of non-linearity' [15] and false solutions [12] in non-linear inverse problems, and limitations arising from the finite number of the so called 'degrees of freedom' of scattered fields [23].

In Section III, a number of alternative rewriting of the scattering equations, that are the contrast source extended Born (CS-EB) model [16]-[17], the family of integral scattering formulations, known as NIE models [18], and the very recent Y0 model [19] are reviewed, commented and compared for the first time. Some comments on the so-called Strong Permittivity Formulation model [24]-[25] (which only can be applied to vector problems) is also given.

In Section IV, turning back to the standard writing of the inverse scattering equations, the 'virtual' scattering experiment framework is recalled [20]-[22]. Finally, in Section V the Y0-NIE-CSI and VE-NIE-CSI hybrid methods are discussed, while in Section VI numerical tests are performed within a non-linear regime and against simulated data. Then, we come to Conclusions.

## II. INVERSE SCATTERING: BASIC EQUATIONS AND ISSUES

Inverse scattering consists in retrieving the electromagnetic properties of unknown targets by properly processing their scattered fields when illuminated by given incident fields $E_i$ [1].

The basic equations underlying the problem are the data and the state equations. The first one relates the scattered field $E_s$ measured on a given measurement curve $\Gamma$ to the electromagnetic properties of the target, which are encoded in the contrast function $\chi$. The state equation, also known as Lippman Schwinger equation, is instead the mathematical expression of the total field (or the currents) induced in the investigation domain $D$ in term of the contrast function $\chi$. This latter is expressed as $\chi = \frac{\varepsilon_x(\underline{r})}{\varepsilon_b(\underline{r})} - 1$, being $\varepsilon_b(\underline{r})$ and $\varepsilon_x(\underline{r})$ the complex permittivities of the background medium and the unknown targets, respectively. By assuming and dropping the time harmonic factor $exp\{j\omega t\}$, the mathematical expressions of data and state equations for the 2D scalar problem are respectively [1]-[3]:

$$E_s(\underline{r}_m, \underline{r}_t) = \int_D G_b(\underline{r}_m, \underline{r}') \chi(\underline{r}) E_t(\underline{r}', \underline{r}_t) dr' = A_e[W(\underline{r}, \underline{r}_t)] \tag{1}$$

and

$$\begin{aligned} W(\underline{r}, \underline{r}_t) &= \chi(\underline{r}) E_i(\underline{r}, \underline{r}_t) \\ &+ \chi(\underline{r}) \int_D G_b(\underline{r}, \underline{r}') \chi(\underline{r}) E_t(\underline{r}', \underline{r}_t) dr' \\ &= \chi(\underline{r}) E_i(\underline{r}, \underline{r}_t) + \chi(\underline{r}) A_i[W(\underline{r}, \underline{r}_t)] \end{aligned} \tag{2}$$

wherein $\underline{r} = (x, y) \in D$, $\underline{r}_t$ and $\underline{r}_m$ are the transmitting and receiving positions on the curve $\Gamma$. Then, $W = \chi E_t$, and $E_t$ are respectively the so-called contrast sources and the total electric fields in D while $A_e$ and $A_i$ are short notations of the external and internal radiation operators, respectively. Finally, $G_b(\underline{r}, \underline{r}') = -\frac{j}{4} k_b^2 H_0^2(k_b|\underline{r} - \underline{r}'|)$ is the Green's function pertaining to the background medium, being $H_0^2$ the zero order and second kind Hankel function and $k_b = \omega\sqrt{\mu_b \varepsilon_b}$ the wavenumber in the host medium.

In order to emphasize the differences with the ones proposed in the following, let us identify the model (1)-(2) as the H0 model.

### A. Issues related to inverse scattering problems

In inverse scattering problems, the solution can be proved to exist in case of noise-free data [23], and, provided a number of hypotheses on the collected data hold true, theoretical uniqueness can be proved in both the tridimensional and bidimensional cases [26],[27].

On the other side, even when the solution exists and it is unique, another element comes into play, so that the problem is anyway ill-posed. In fact, one can easily prove that the solution does not depend continuously on data, which comes essentially from the fact that the operator $A_e$ is compact so that it cannot be inverted. Moreover, another consequence of the compactness of $A_e$ is that there is no hope of extending at will the number of independent scattering experiments, but only a finite number of receivers and transmitters actually bring independent information. In fact, care has to be taken in choosing the positions ($\underline{r}_t$ and $\underline{r}_m$) and the number ($N_T$ and $N_R$) of transmitting and receiving probes, in such a way to collect all the available information in a non-redundant fashion. This can be efficiently done by adopting the measurement strategies proposed in [23], wherein a Nyquist criterion is essentially suggested for the case at hand.

Unfortunately, the inverse scattering problem is also non-linear, because of the dependence of $W$ or $E_t$ on the unknown contrast function $\chi$ [1],[3]. In a first instance, non-linearity can be attributed to the mutual (or even self) interactions amongst the different parts of the scatterer. A generalized solution of inverse scattering is usually looked for by minimizing a suitable cost functional, which takes into account misfits in both the data and state equations. Due to the non-linearity of the underlying problem, this cost functional is a non-quadratic one, so that it may have many local minima which are 'false solutions' of the problem [12]. The more the problem departs from a linear one, the larger the chance of incurring into a so called 'false solution'. In fact, if the initial guess does not belong to the attraction region of the actual solution, the minimization scheme could be trapped into local minima completely different from the actual ground truth.

As discussed in [12] a possible key to avoid the occurrence of false solutions is to have a sufficiently large ratio amongst the number of independent information and number of unknowns. Unfortunately (see above) the number of independent data cannot be increased at will because of the limited number of degrees of freedom of scattered fields [23], so that additional care and tools are needed, as discussed in the following.

### B. Degree of non-linearity

In order to quantify the degree of non-linearity (DNL) of the relationship between the unknown permittivity profile and scattered field, a useful and effective key is reasoning in terms of the norm of the operator $\chi A_i$. In fact, if this norm is lower than 1, the inverse operator formally solving eq. (2) can be expanded into a Neumann series, that is:

$$(I - \chi A_i)^{-1} = I + \chi A_i + (\chi A_i)^2 + \cdots + (\chi A_i)^n + \cdots \tag{3}$$

Notably, the condition $\|\chi A_i\| < 1$ is just a sufficient condition for the convergence of the above series, so that series (3) can eventually converge even if the above condition is not satisfied.

By evaluating the norm of $\chi A_i$, one can also foresee what is the number of series terms to be adopted in order to achieve a given approximation accuracy [15]. Moreover, one can understand that the overall DNL and, hence, the complexity and difficulty of the inverse scattering problem at hand, increases with the norm of the operator $XA_i$.

Then, understanding the factors affecting $\|\chi A_i\|$ is key to understand and eventually reduce the occurrence of false solutions. In this respect, let us consider that, by applying the Schwarz's inequality, an upper bound to the norm of interest can be obtained as:

$$\|\chi A_i\| < \|\chi\|\|A_i\| \tag{4}$$

In such a way, the role played by the internal radiation operator $A_i$, which only depends on the kernel of the integral operator and on the considered domain, is separated by the one played by contrast function $\chi$. Hence, for a fixed contrast function, the non-linearity of the problem depends on the structure of internal radiation operator $A_i$ in the state equation. On the other hand, for a fixed structure of the radiation operator in the state equation, the DNL only depends on the norm of $\chi$.

### III. EVALUATING THE DEGREE OF NON-LINEARITY OF DIFFERENT REWRITINGS OF THE SCATTERING EQUATIONS

In the following, three different rewritings of the scattering equations are reviewed, that are the contrast source extended Born CS-EB model (and the related family of equations) [16],[17], the family of new integral equations, known as NIE model [18], and the very recent Y0 model [19].

#### A. CS-EB model

By exploiting the Green's function peaked behavior in the presence of losses in the background medium and by using similar concepts as the ones adopted for deriving the extended Born (EB) approximation of [28], in [16],[17] a rewriting of the traditional H0 model was proposed. This approximation-free model hinges on the extraction of some dominant contribution to the integral in (2) (as written in terms of the contrast source). As a matter of fact, after simple manipulations (see [16],[17] for more details) one can rewrite the state equation (2) as follows:

$$W(\underline{r},\underline{r}_s) = p(\underline{r})E_i(\underline{r},\underline{r}_s) + p(\underline{r})A_i^{CS-EB}[W(\underline{r},\underline{r}_s)] \tag{5}$$

wherein:

-   $A_i^{CS-EB}[\,\cdot\,] = A_i[\,\cdot\,] - f_D(\underline{r})I$; (6.a)

-   $f_D(\underline{r}) = \int_D G_b(\underline{r},\underline{r}')\,dr'$; (6.b)

-   $p(\underline{r}) = \frac{\chi(\underline{r})}{1-\chi(\underline{r})f_D(\underline{r})}$. (6.c)

being $I$ the identity operator. Equation (5) together with (1) identifies the CS-EB model, wherein the fundamental quantities are the contrast sources $W$ and the auxiliary function $p$, which is the one now encoding the properties of the scatterer. Note the structure of the state equation (5) is identical to the one of the H0 model, where the integral internal operator $A_i$ and the function $\chi$ have been now replaced by $A_i^{CS-EB}$ and by the auxiliary variable $p$, respectively.

The function $f_D(\underline{r})$ can be evaluated in a closed form in many cases of practical interest. For example, in case of a circular cylinder of radius $a$, its expression is given as [17]:

$$f_D(\underline{r}) = -\frac{j\pi k_b a}{2} H_1^{(2)}(k_b a) J_0(k_b|\underline{r}|) - 1 \tag{7}$$

wherein $H_1^{(2)}$ is the Hankel function of first order and second kind while $J_0$ is the Bessel function of zeroth order.

If the radius $a$ is assumed as a parameter, apart from the domain D at hand, the CS-EB model can be interpreted as a family of equivalent equations. So, for a given domain D wherein the operator $A_i[\,\cdot\,]$ is evaluated, a proper choice of $a$ can be eventually exploited in order to optimize the performance of inversion procedures. By exploiting such a further degree of freedom, it is worth to note that the CS-EB model, although introduced for the case of lossy scenarios, is indeed of interest also for the lossless case (see for more details [17]).

A comparison between the H0 model (1)-(2) and the CS-EB model (1)-(5) can be performed in terms of DNL. By virtue of the inequality (4), one can compare the right-hand factors of (4) model and the corresponding ones regarding the CS-EB model, that is:

$$\|pA_i^{CS-EB}\| < \|p\|\|A_i^{CS-EB}\| \tag{8}$$

In particular, paper [17] proves that $\|A_i^{CS-EB}\| \approx \|A_i\|$ (see also subsection III.D for more details) and, hence, the CS-EB and H0 models are almost equivalent as far as the involved integral operators are concerned. Then, one can argue that the CS-EB model has a reduced DNL with respect to H0 model when $\|p\| < \|\chi\|$, which allows to establish 'convenience maps' for using one of the two models for given categories of scatterers [17]. For instance, as explained in [17], provided $a$ is properly chosen, the CS-EB model can be conveniently adopted in case of completely lossless scenario and contrast having a positive real part.

As for the model described in the following subsection, the contrast function is embedded into an auxiliary unknown function, so that some care has to be taken into the additional step requiring the extraction of the actual unknown function from the auxiliary one.

#### B. NIE model

In the same line of reasoning of CS-EB model and motivated by the contraction integral equation [29] in tackling forward problems with strong conductivities, in [18],[30] a family of new integral equations (NIE) is proposed, which are also derived by rewriting in a different fashion the standard state equation (2). By extracting a local effect of the induced currents [18] and introducing a new convenient auxiliary unknown the amount (and the effects) of non-linearity can be alleviated. In particular, in the NIE model the state equation is rewritten[1] (again, without any approximation) as follows:

$$\beta(\underline{r})W(\underline{r},\underline{r}_t) = R(\underline{r})\beta(\underline{r})W(\underline{r},\underline{r}_t) \\ + R(\underline{r})\Big[E_i(\underline{r},\underline{r}_t) + A_i\big[W(\underline{r},\underline{r}_t)\big]\Big] \tag{9}$$

where:

$$R(\underline{r}) = \frac{\beta(\underline{r})\chi(\underline{r})}{\beta(\underline{r})\chi(\underline{r}) + 1} \tag{10}$$

is the modified contrast function and $\beta$ is such that the denominator of $R(\underline{r})$ is different from zero.

The term $R(\underline{r})\beta(\underline{r})W(\underline{r},\underline{r}_t)$ represents the local effect of the induced currents. The equation (9) can be rewritten such that it exhibits the same structure of the state equation (2), i.e.:

$$\beta(\underline{r})W(\underline{r},\underline{r}_t) = R(\underline{r})E_i(\underline{r},\underline{r}_t) + R(\underline{r})A_i^{NIE}\big[\beta(\underline{r})W(\underline{r},\underline{r}_t)\big] \tag{11}$$

wherein:

$$A_i^{NIE}[\cdot] = I + \frac{A_i[\cdot]}{\beta(\underline{r})} \tag{12}$$

With respect to equation (2), the integral internal operator $A_i$, the induced currents and the function $\chi$ have been now replaced by $A_i^{NIE}$, $\beta W$ and $R$, respectively.

Note that different functions $\beta(\underline{r})$ lead to different NIE integral equations. By properly selecting $\beta(\underline{r})$, the local term $R(\underline{r})\beta(\underline{r})W(\underline{r},\underline{r}_t)$ can be predominant and it is possible to reduce the associated DNL with respect to the H0 model (1)-(2). As the structure of (11) is still the same as for the H0 model and for the CS-EB model, the DNL of the NIE model can be analyzed according to the norm of $RA_i^{NIE}$. Then, as

$$\|RA_i^{NIE}\| < \|R\|\|A_i^{NIE}\| \tag{13}$$

one can separately evaluate the role played by the new auxiliary unknown R and by the involved operator by means of $\|R\|$ and $\|A_i^{NIE}\|$, respectively. If $\beta(\underline{r})\,\chi(\underline{r})$ has a positive real part, the norm $\|R\|$ is always less than 1. As a consequence, in case of high contrast targets with positive real part and non-negative imaginary part of $\chi$, this condition holds true with $\beta(\underline{r})$ having a positive real part and a non-positive imaginary part [18]. On the other side, this is just a part of the story, as the norm of $A_i^{NIE}$ has to be checked as well (see the Subsection III.D below for more details).

Note that, as $\beta$ increases, the auxiliary contrast tends to 1 and plays the role of a support indicator. Unfortunately, this has an impact on the step leading from the auxiliary unknown to the actual contrast. In fact, very large $\beta$ implies a kind of binary behavior of $R(\underline{r})$, so that one would have to extract a generically varying function from a binary one.

## C. Y0 model

The very recent Y0 model [19] is based on a convenient decomposition of the pertaining Green's function, which allows to extract the main contribution of the radiating currents to the total field from the internal integral operator. In detail, the Y0 model hinges on the following rewriting of the state equation:

$$W(\underline{r},\underline{r}_t) = \chi(\underline{r})\hat{E}_i(\underline{r},\underline{r}_t) + \chi(\underline{r})A_i^{Y_0}\big[W(\underline{r},\underline{r}_t)\big] \tag{14}$$

wherein:

- $\hat{E}_i(\underline{r},\underline{r}_t) = E_i(\underline{r},\underline{r}_t) - j\frac{k_b^2}{4}F_{J_0}(\underline{r},\underline{r}_t)$ (15.a)

- $F_{J_0} = \int_D J_0(k_b|\underline{r} - \underline{r}'|)\,W(\underline{r}',\underline{r}_t)d\underline{r}'$ (15.b)

- $A_i^{Y_0}[\cdot] = -\frac{k_b^2}{4}\int_D Y_0(k_b|\underline{r} - \underline{r}'|)\,[\cdot]d\underline{r}'$ (15.c)

As discussed in [19], $-j\frac{k_b^2}{4}F_{J_0}{}^v$ can be understood as the contribution to the total field inside D by the main spectral component of the radiating currents, which are indeed peaked in the spectral domain along the circle of radius $k_b$. Notably, this contribution can be easily computed from the data (see [19] for more details).

Together with (1), the state equation (14) identifies the Y0 model, wherein the unknowns are again the induced sources $W$ and the contrast function $\chi$. However, with respect to the H0 model, the integral internal operator $A_i$ has been now replaced by $A_i^{Y_0}$ and the incident field $E_i$ by a new known 'modified incident' field $\hat{E}_i$.

Again, by virtue of the inequality (4), one can separately analyze the roles of the contrast profile and of the relevant integral operator with respect to the DNL of the problem at hand. In particular, one can again compare the quantities at the right-hand side of (4) and of the corresponding inequality which holds true for the Y0 model, i.e.,

$$\|\chi A_i^{Y_0}\| < \|\chi\|\|A_i^{Y_0}\| \tag{16}$$

As the factor $\|\chi\|$ is the same in the two inequalities (4) and (16), the DNL of the two formulations can be eventually compared by just evaluating $\|A_i^{Y_0}\|$ and $\|A_i\|$. As shown in [19] and discussed in the following Subsection, in case of homogeneous and lossless background, the Y0 model exhibits a lower DNL.

## D. Comparisons of the three models

All the above models are based on a convenient rewriting of the state equation which lead to the same structure, that is:

$$W^{MOD}(\underline{r},\underline{r}_t) = \chi^{MOD}(\underline{r})E_i^{MOD}(\underline{r},\underline{r}_t) \\ + \chi^{MOD}(\underline{r})A_i^{MOD}\big[W^{MOD}(\underline{r},\underline{r}_t)\big] \tag{17}$$

---

[1] In short, eq. (9) can be derived by adding the term $\beta\chi W$ at both sides of eq. (2) and, then, by normalizing the resulting equation to the term $\beta\chi + 1$.

wherein, depending on the adopted model, the function $\chi^{MOD}$, $W^{MOD}$, $E_i^{MOD}$ and the operator $A_i^{MOD}$ assume the proper meaning, that are:

$$W^{MOD} = \begin{cases} W & in\ CS-EB\ model \\ \beta W & in\ NIE\ model \\ W & in\ Y0\ model \end{cases}$$
(18.a)

$$E_i^{MOD} = \begin{cases} E_i & in\ CS-EB\ model \\ E_i & in\ NIE\ model \\ E_i - j\dfrac{k_b^{\ 2}}{4}F_{J_0}^{\ v} & in\ Y0\ model \end{cases}$$
(18.b)

$$\chi^{MOD} = \begin{cases} p & in\ CS-EB\ model \\ R & in\ NIE\ model \\ \chi & in\ Y0\ model \end{cases}$$
(18.c)

$$A_i^{MOD}[\ \cdot\ ] = \begin{cases} A_i[\ \cdot\ ] - f_D I & in\ CS-EB\ model \\ \dfrac{A_i[\ \cdot\ ]}{\beta} + I & in\ NIE\ model \\ -\dfrac{k_b^{\ 2}}{4}\displaystyle\int_D Y_0(k_b|\underline{r}-\underline{r}'|)[\ \cdot\ ]d\underline{r}' & in\ Y0\ model \end{cases}$$
(18.d)

Some comments are now in order.

In the three models, the internal operator is modified in order to hopefully reduce the corresponding norm, while the unknown function of (17) keeps unchanged with respect to H0 model only in the Y0 formulation. On the other hand, the 'contrast sources' are re-defined (actually, scaled) in the NIE model, thus implying a trivial formal modification of the external radiation operator.

The NIE model, which can be seen as a generalization of the CS-EB model, includes the latter as a special case. Indeed, if $\beta(\underline{r}) = -f_D(\underline{r})$ the NIE model for the state equation becomes essentially identical to the CS-EB model. As already stated, care has to be taken in selecting both $\beta(\underline{r})$ and $f_D(\underline{r})$, respectively, as they could significantly imperil the reverse mapping from modified contrast function to physical contrast $\chi$. As also discussed at the end of Subsection III.B, the lower the dynamic of $p$ and $R$, the more difficult the extraction electromagnetic properties of the targets.

Differently from CS-EB model, wherein $f_D(\underline{r})$ is given as a function of $\underline{r}$ [17], in [18] a constant value of $\beta(\underline{r})$ is selected by considering that all small cells in the domain of interest are equivalent in terms of the possibility of being a scatterer or not. As discussed in [18], this can be a reasonable choice, whereas the optimal selection of $\beta(\underline{r})$ remains anyway an open problem. Note that, the choice of a constant $\beta$ also allows to transform a piecewise constant contrast profile $\chi(\underline{r})$ into piecewise constant behavior of $R(\underline{r})$, which can be an advantage in regularized inversions based on total variation and/or compressive sensing (see for instance [31],[32]).

Differently from CS-EB and NIE, in the Y0 model the geometrical and electromagnetic properties of the targets and the induced currents are still encoded in the contrast function $\chi$ and $W$, respectively. As a consequence, there is no need to adopt further procedures to extract the target features. Finally, it is worth to note that while CS-EB and NIE models can also be used for the solution of the forward problem, the Y0 model is instead a data driven model, wherein the 'modified incident' field includes a term which is (easily) computed from the data.

In order to quantitative compare the DNL of the three models, a numerical analysis has been performed by following the same reasoning as in [17],[19] and by considering a circular domain D of radius $r_D$. Note that, to the best of authors knowledge, this is the first time that the norm of $\|A_i^{NIE}\|$ is analyzed as well as a comparison is performed among the three models. The plots of the norm of the relevant operators as a function of $r_D/\lambda$, where $\lambda$ is the wavelength in the background medium, are shown in Figure 1. By the sake of simplicity, just the case of lossless background is considered. As can be seen, whatever the model, the norms are monotonically increasing functions of $r_D/\lambda$.

Interestingly, one can notice that $\|A_i^{Y_0}\|$ is always lower than $\|A_i\|$, as also shown in [19]. As such, for a fixed scattering problem, the Y0 model exhibits a lower DNL with respect to the H0 model and, hence, a smaller occurrence of false solutions.

As far as $\|A_i^{NIE}\|$, as expected from its expression, it starts from 1, which is not a very favorable condition in view of the condition for the convergence of the corresponding Neumann expansion (3). On the other side, this is just one of the factors at the right-hand side of (13). Moreover, the condition on the norm of the overall operator entering the state equation is just a sufficient condition, so that the series can converge anyway[2].

As can be seen from figure 1(a), a key role is played by the parameter $\beta$. By following [18], $\beta$ has been selected as a real constant value belonging to the interval [0.5,6]. As can be seen, in case of $\beta > 1$ the NIE model exhibits a lower norm of $\|A_i^{MOD}\|$, with respect to H0 model.

Finally, in case of CS-EB model, as already recalled in Section III.A, the two norms $\|A_i\|$ and $\|A_i^{CS-EB}\|$ are quite similar, (except for the case of $r_D < 0.5\ \lambda$), so that actual choice amongst CS-EB and the original and more widespread model depends on a comparison amongst the expected values of $p$ and $\chi$.

As these norms are just a part of the story, the corresponding norm of the auxiliary contrast variables, whose maximum also enters in the evaluation of the bound of $\|\chi^{MOD}A_i^{MOD}\|$, are also reported in Figures 1(e)-(f). These norms are evaluated as $max|\chi^{MOD}|$ according to [17]. In particular, the norms of the auxiliary contrast functions $R(\underline{r})$ and $p(\underline{r})$ in case of homogeneous circular cylinder of radius $1\lambda$ and different $\chi$ values are shown by changing both $\beta$ in eq. (10) and $a$ in (7). As can be seen, in case of NIE model, $\|R\|$ is monotonically increasing functions of $\beta$ and $\chi$. In particular, if $\beta$ increases, $\|R\|$ tends quickly to 1. On the other hand, in case of CS-EB model, the $\|p\|$ is an oscillating function of $a$. Also, some points can exist wherein the term $1 - \chi(\underline{r})f_D(\underline{r}) \ll 1$ and are cause of the peaks in Figure 1(f).

---

[2] In the inverse problem, in one of the authors experience a rule of thumb for successful (false solutions free) inversion (for full aspect experiments) is $\|\chi^{MOD}\|\|A_i^{MOD}\| < 2$.

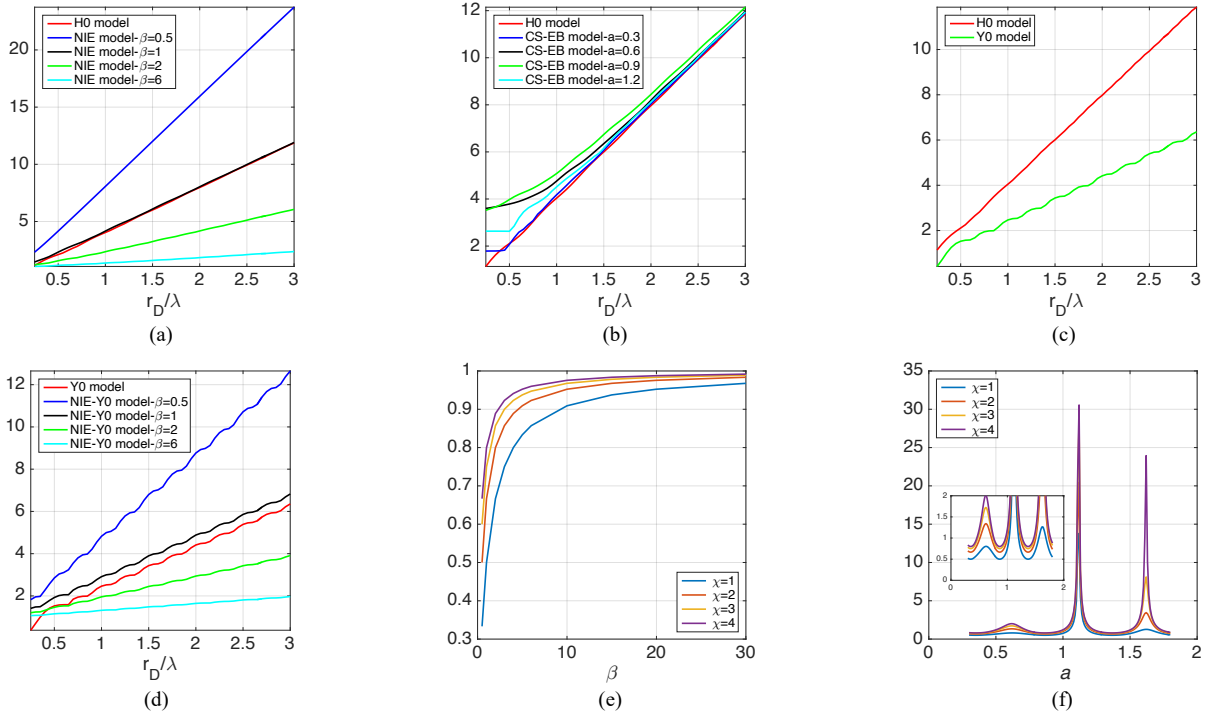

Fig. 1. Evaluating the degree of non-linearity of different rewritings. Norms of the internal radiation operations. (a) $\|A_i\|$ versus $\|A_i^{NIE}\|$ in case of $\beta = 0.5$, $\beta = 1$, $\beta = 2$ and $\beta = 6$. (b) $\|A_i\|$ versus $\|A_i^{CS-EB}\|$ in case of $a = 0.3$, $a = 0.6$, $a = 0.9$ and $a = 1.2$. (c) $\|A_i\|$ versus $\|A_i^{Y0}\|$. (d) $\|A_i^{Y0}\|$ versus $\|A_i^{NIE-Y0}\|$ in case of $\beta = 0.5$, $\beta = 1$, $\beta = 2$ and $\beta = 6$. Norms of the auxiliary contrast functions in case of homogeneous circular cylinder of radius $1\lambda$ and different $\chi$ values. (e) $\|R\|$ versus $\beta$. (f) $\|p\|$ versus $a$. Note that in (f) a zoom in on a portion of a curve is displayed.

Note that even in case of large (positive) contrast, and large $\beta$, the maximum norm of $R$ is less than 1 (and of the corresponding $\|\chi\|$). Such a circumstance, and results in figure 1(a), justifies the remarkable performance of NIE in inverse scattering. As already argued, a price is somehow paid in an increased difficulty is the final mapping from the modified contrast function to the actual physical contrast $\chi$. In this respect, let us also note that very large $\beta$ implies that the state equation (9) becomes an identity, where the internal radiation operator $A_i$ in eq. (10) (and hence the overall state equation) just disappears (see also the footnote at the previous page).

The given graphs for the three models and the corresponding expressions for the (auxiliary) variables can give guidelines in the choice of the most convenient model to be used in the inversion for given classes of expected contrast profiles.

Another interesting direction to be pursued is the possible hybridization of the different models, which is partially addressed in Section V below.

### E. Strong Permittivity Fluctuation

Even if we focus our attention on 2D scalar TM inverse scattering problem, it is worth to note that also the so-called Strong Permittivity Fluctuation (SPF) model [24], [25] (which is applied to vector problems) can be interpreted as a rewriting of relevant scattering equations. In particular, according to SPF model, the singularity of the dyadic Green's function is extracted and the TE state equation can be rewritten in an alternative fashion as:

$$\underline{W}(\underline{r}, \underline{r}_t) = \chi^{MOD}(\underline{r})\underline{E}_i(\underline{r}, \underline{r}_t) + \chi^{MOD}(\underline{r})\underline{\underline{A}}_i^{PV}[\underline{W}(\underline{r}, \underline{r}_t)] \tag{19}$$

wherein:

- $\chi^{MOD}(\underline{r}) = \frac{2\chi(\underline{r})}{2 + \chi(\underline{r})}$; \hfill (20.a)

- $\underline{\underline{A}}_i^{PV}[\,\cdot\,] = P.V. \cdot \int_D \underline{\underline{G}}_b(\underline{r}, \underline{r}')\,[\,\cdot\,]d\underline{r}'$ \hfill (20.b)

with P.V. denoting the principal value integral.

The new TE state equation (19) has the same structure of the previous ones but for the vector nature. The target properties are now encoded in the variable $\chi^{MOD}$ of eq. (20.a), while the internal radiation operators are now replaced with the corresponding principal value integral $\underline{\underline{A}}_i^{PV}$ in eq. (20.b).

As the CS-EB and NIE ones, the SPF model can be adopted for the solution of both forward and inverse problems, and provided the corresponding norms are evaluated the expressions above also may serve to compare the degree of non-linearity of the TE equations as compared to the TM ones.

## IV. Rewriting experimental data in terms of Virtual Experiments

In Section III we have discussed how different rewritings of the relevant state equation allow to understand and possibly reduce the DNL. In this section we focus on possible rewritings of the equations describing the different scattering experiments through proper superpositions of the data. In a nutshell, the 'virtual experiments' (VE) framework [20]-[22] acts on the data in order to condition the (implicit) scattering phenomenon and enforce peculiar and hopefully useful properties of the unknown contrast sources (or total fields) induced in the imaging domain D.

By assuming linear constitutive relationships for a fixed contrast function, the scattering phenomenon is linear with respect to the primary sources. Then, a superposition of the incident fields coming from the $N_T$ different transmitters located in $\underline{r}_t$, with known coefficient $\alpha$,:

$$\mathcal{E}_i(\underline{r}) = \sum_{t=1}^{N_T} \alpha(\underline{r}_t) \, E_i(\underline{r}, \underline{r}_t)$$

(21)

gives rise to a scattered field and contrast source which are nothing but the superposition with the same coefficients of the corresponding scattered fields and contrast sources, respectively, so that:

$$\mathcal{W}(\underline{r}) = \sum_{v=1}^{N_T} \alpha(\underline{r}_t) \, W(\underline{r}, \underline{r}_t)$$

(22.a)

$$\mathcal{E}_s(\underline{r}_m) = \sum_{v=1}^{N_T} \alpha(\underline{r}_t) \, E_s(\underline{r}_m, \underline{r}_t)$$

(22.b)

According to (21) and (22), the original data can be reorganized in a possibly more convenient way by means of a linear superposition of the incident fields. However, the amount of information carried by VE cannot exceed that of the original ones, and some information could be actually lost if they are not properly designed.

By acting on the coefficients $\alpha$, one can perform several re-arrangements of the original experiments and thus build a set of virtual experiments [20]-[22]. These new experiments do not require additional measurements and are derived from a-posteriori software procedures. Also, no a priori information on the contrast function is needed to generate VE.

In the VE framework, the H0 model can be recast as:

$$\mathcal{W}(\underline{r}) = \chi(\underline{r})\mathcal{E}_i(\underline{r}) + \chi(\underline{r})A_i[\mathcal{W}(\underline{r})]$$

(23)

$$\mathcal{E}_s(\underline{r}_m, \underline{r}_t) = A_e[\mathcal{W}(\underline{r})]$$

(24)

With respect to model (1)-(2), the above equations have the same structure. Moreover, the integral operators as well as the function encoding the target proprieties are the same. However, the electromagnetic variables, that are the scattered data, the incident field and the induced currents are now replaced with $\mathcal{E}_s$, $\mathcal{E}_i$ and $\mathcal{W}$.

Amongst the different possibilities and considering the case when the scatterers are hosted in a homogeneous background, a convenient VE design is represented by the rearrangement of the original data in such a way to enforce a set of contrast sources focused on a set of different 'pivot points' $\underline{r}_p$.

Provided some hypotheses (for instance, on the shape of the target) hold true [20]-[22], such request implies a localization of the scattering phenomenon so that the mutual interactions, which are one of the causes of the non-linearity of the problem, are minimized.

The VE framework has opened the way to innovative, convenient and effective solution procedures. For instance, a new data-driven linear approximation has been developed, which has an extended validity range as compared to the traditional Born approximation [20]. Moreover, effective non-linear inversion methods based on algebraic inversion formulas [21] or peculiar forms of regularization [22] have been proposed. Other interesting realizations of the basic 'Virtual Experiments' idea imply 'Distorted Virtual Experiments' [33] and 'Distorted Iterated Virtual Experiments' [34]. Another example of non-iterative linear inversion method based on VE is proposed in [35].

In the following, we briefly recall one of them, namely the iterative approach in [22] as it is going to be used it in Section V.

### A. Regularized VE - CSI

Let us consider the contrast source inversion (CSI) method [10], where the inverse problem is cast as the minimization of a cost functional accounting for both the misfit in the data and state equations. Such a functional depends on both the unknown contrast and the auxiliary unknown, which is the contrast source therein induced.

In the VE framework, if $\underline{r}_p^{(1)}, \ldots, \underline{r}_p^{(P)}$ is a set of $P$ pivot points inside the target support, the first step amounts to design virtual experiments such that a circular symmetry of the contrast sources (around the pivot point) is enforced (see [22] for more details). Then, it is possible to recast the standard CSI scheme in terms of the thus designed VE as the minimization of [22]:

$$\Phi_{VE}(\chi, \mathcal{W}^{(p)}) = \sum_{p=1}^{P} \frac{\left\| \mathcal{W}^{(p)} - \chi\mathcal{E}_i^{(p)} - \chi A_i[\mathcal{W}^{(p)}] \right\|_{\mathrm{D}}^2}{\left\| \mathcal{E}_i^{(p)} \right\|_{\mathrm{D}}^2} +$$
$$+ \sum_{p=1}^{P} \frac{\left\| \mathcal{E}_s^{(p)} - A_e[\mathcal{W}^{(p)}] \right\|_{\Gamma}^2}{\left\| \mathcal{E}_s^{(p)} \right\|_{\Gamma}^2} + \Phi_{\mathcal{W}}(\mathcal{W}^{(1)}, \ldots, \mathcal{W}^{(P)})$$

(25)

With respect to the standard CSI method, the cost functional is equipped with the additional regularizing term $\Phi_{\mathcal{W}}$. In particular, given the VE design criterion which has been used, the additional term is given by:

$$\Phi_W\left(\mathcal{W}^{(p)}\right) = \sum_{p=1}^{P} \tau_p \left\|\frac{\partial \mathcal{W}^{(p)}}{\partial \phi_p}\right\|_{\Omega}^2, \qquad \underline{r} \in \mathcal{I}_R\left(\underline{r}_p^{(p)}\right)$$

(26)

where $\phi_p$ is the angular coordinate of a local polar reference system centered in $\underline{r}_p^{(p)}$ which spans the circular neighborhood $\mathcal{I}_R\left(\underline{r}_p^{(p)}\right)$, and $\{\tau_p\}_1^P$ are non-negative parameters controlling the relative weight of such a regularization term.

The penalty term (26) is a way to enforce the expected contrast sources properties by minimizing the angular variation of each $\mathcal{W}^{(p)}$ around the pertaining pivot point $r_p$. Indeed, the functional $\Phi_W$ encourages the research of circularly symmetric sources, while angularly varying contrast sources are penalized.

Note that, differently from most of the regularization schemes adopted in conjunction with CSI method (see for instance [10]), the non-linearity of the problem is herein tackled by acting on the contrast sources rather than by relying on a priori information on the contrast function.

## V. POSSIBLE HYBRIDIZATIONS OF THE ABOVE CONCEPTS

All the above (equations or data) rewritings have proved their usefulness. Interestingly, one can indeed combine two (or even more) different methods, thus hopefully giving rise to even more powerful inversion methods. Notably, such a simple consideration opens the way to a wealth of new possible solution methods in inverse scattering.

In this section, we introduce two new inversion procedures, whose interest and effectiveness are then proved in the numerical experiments in Section VI.

### A. Y0 − NIE model

As a first possible hybridization, one can consider to sum up the advantages offered by the Y0 model and the NIE family of possible formulations. As discussed above, the Y0 model can offer the advantage of reducing the DNL of the problem by extracting the contribution of the dominant part of the radiating currents. On the other hand, by means of a re-definition of both the contrast function and of the operator entering the state equation, NIE allow for a (different) way of reducing the degree of non-linearity.

As already argued, in NIE the higher $\beta$, the higher the norm of R, and the harder the inversion procedure to retrieve the electromagnetic properties of the targets from the auxiliary variable. Hence, one can reasonably combine the advantages of NIE and Y0 models to come to a hybrid 'Y0-NIE' model. In particular, in such model the state equation can be recast as follows,

$$\beta(\underline{r})W(\underline{r},\underline{r}_t) = R(\underline{r})\hat{E}_i(\underline{r},\underline{r}_t) + R(\underline{r})A_i^{Y0-NIE}[\beta(\underline{r})W(\underline{r},\underline{r}_t)]$$

(27)

wherein:

$$A_i^{Y0-NIE}[\,\cdot\,] = \frac{A_i^{Y0}[\,\cdot\,]}{\beta(\underline{r})} + I$$

(28)

In eq. (27) not only the local effect of the induced currents (as discussed in [18]) but also the contribution of the radiating currents is isolated (as discussed in [19]), thus further reducing the DNL of the inverse scattering problem.

In order to quantitative evaluate the advantages, a numerical analysis is performed as in Section III.D by considering a circular domain D of radius $r_D$. The plots of the norm of the Y0 and Y0-NIE operators as a function of $r_D/\lambda$ are shown in Figure 1(d). As expected, the norm of the relevant operator is further reduced with respect to the norms in Figures 1(a) and 1(d).

### B. VE − NIE model

A different hybridization can be obtained by jointly rewriting the scattering equations and the equations regarding the different experiments.

In particular, let us develop here a hybridization of the NIE model(s) with the VE framework in order to hopefully benefit from both concepts. In such a hybridization, the state equation can be recast (for each pivot point $\underline{r}_p^{(p)}$) as follows:

$$\beta(\underline{r})\mathcal{W}^{(p)}(\underline{r}) = R(\underline{r})\mathcal{E}_i^{(p)}(\underline{r}) + R(\underline{r})A_i^{NIE}[\beta(\underline{r})\mathcal{W}^{(p)}(\underline{r})]$$

(29)

which, together with (24), identifies the VE-NIE model. Note that in eq. (29) the local effect of the induced currents is isolated by means of the rewriting of equation (1) according to [18]. Moreover, this local effect is emphasized by designing circularly symmetric VE [20]-[22].

The model could be exploited and valorized in conjunction with all the different methods introduced in the VE framework [20]-[22]. Herein, let us focus our attention on the regularized VE − CSI of Section IV.A [22]. Then, by a straightforward correspondence the inverse scattering problem can be recast as the minimization of the functional:

$$\Phi_{VE-NIE} = \sum_{p=1}^{P} \frac{\left\|\beta\mathcal{W}^{(p)} - R\mathcal{E}_i^{(p)} - RA_i^{NIE}\left[\beta\mathcal{W}^{(p)}\right]\right\|_D^2}{\left\|\mathcal{E}_i^{(p)}\right\|_D^2} +$$
$$+ \sum_{p=1}^{P} \frac{\left\|\mathcal{E}_s^{(p)} - A_e[\mathcal{W}^{(p)}]\right\|_\Gamma^2}{\left\|\mathcal{E}_s^{(p)}\right\|_\Gamma^2} + \Phi_W\left(\mathcal{W}^{(1)}, \dots, \mathcal{W}^{(P)}\right)$$

(30)

which is equipped by the same penalty term (26) used before, as we are still looking for circularly symmetric currents.

## VI. NUMERICAL ASSESSMENT

Performance of the hybrid models proposed in Section V have been tested within a non-linear regime. In particular, the CSI method has been adopted to solve the relevant inverse scattering equations. In the following, we referred to these strategies as Y0-NIE-CSI and VE-NIE-CSI, respectively.

As we are just interested herein in focusing on the effects of combining different methods, we are going to exploit here very simple (or no) regularization techniques. In fact, the additional equipping of the proposed techniques with more sophisticated regularization techniques, which is very interesting, would probably confuse the effects of the different actors.

As a matter of fact, in order to defeat ill-posedness, herein the unknown contrast profile is just projected onto a finite number of spatial Fourier harmonics [12]. In particular, the number $N_{arm}$ of sought unknown Fourier coefficients is set equal to the degrees of freedom of scattered fields corresponding to the given region of interest [23].

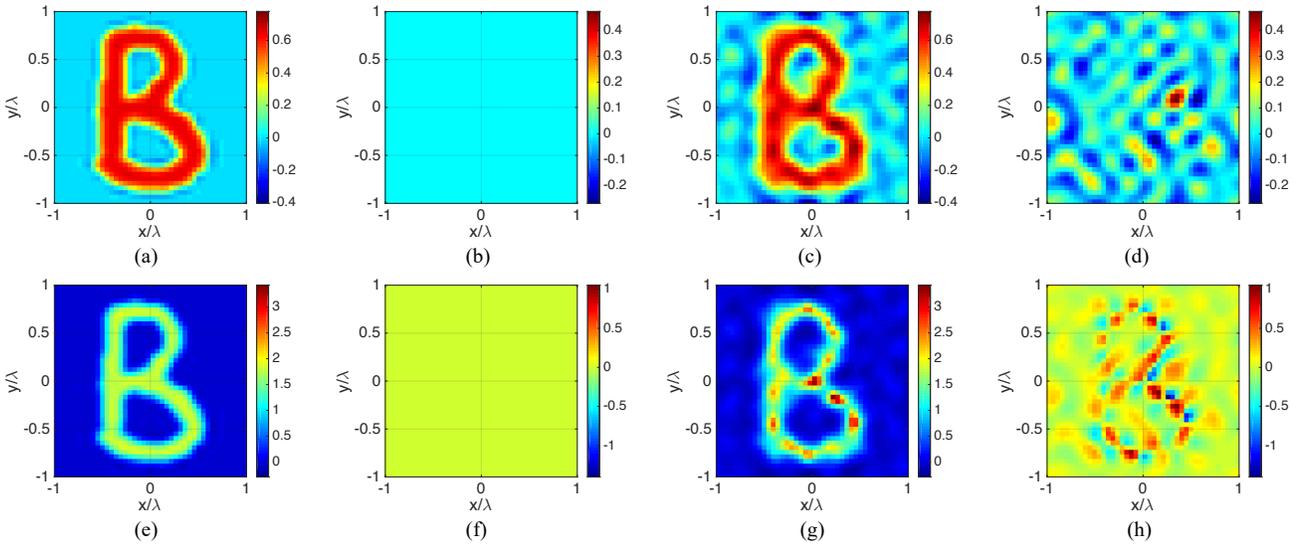

Fig. 2. Y0-NIE model. Assessment against numerical data: B target $\chi$=1.9. Real (a) and imaginary (b) parts of the R reference profile. Real (c) and imaginary (d) parts of the retrieved R function (NMSE=0.17). Real (e) and imaginary (f) parts of the reference contrast profile. Real (g) and imaginary (h) parts of the retrieved contrast function (NMSE=0.18).

Also note that in both approaches, once the profile $R$ is estimated, the properties of the encoded target are simply determined by considering a point-wise inversion of eq. (10). This inversion is troublesome as possible errors in $R$ can be amplified in retrieving $\chi$. Of course, more robust inversion procedures can be adopted by eventually enforcing some expected properties of the targets, but description and adoption of these techniques would further weigh this already long contribution down.

The test target, depicted in Figures 2 and 3, belongs to the MNIST data set [36], which contains both handwritten numerals and letters and is a standard benchmark for machine learning based approaches [14],[37],[38]. Before any numerical calculation, all above equations have been discretized using the method of moments [39]. We choose a square domain of side $L = 2\lambda_b$ discretized into $N_x \times N_y$ small cells, wherein the induced currents and the electromagnetic properties are assumed to be constant, being $N_x = 42$ and $N_y = 42$ the number of cells along the $x$ and $y$ direction.

The target is probed by means of $N_T = N_R = 18$ receivers and transmitters [23], modelled as line sources located on a circumference $\Gamma$ of radius R = 3.75 $\lambda_b$. The scattered field data, simulated at the frequency of 300 MHz by means of a full wave in-house forward solver based on MoM, have been corrupted with a random Gaussian noise with a SNR = 30dB.

The normalized mean square errors between the retrieved contrast function $\tilde{\chi}$ and the actual one $\chi$, defined as:

$$NMSE = \frac{\|\chi - \tilde{\chi}\|^2}{\|\chi\|^2}$$

(31)

have been evaluated in order to quantitatively evaluate the obtained different performance. In all cases, the back-propagation solution has been used as starting guess of the iterative minimization.

### A. Y0 − NIE − CSI results

Figure 2 shows the Y0-NIE-CSI results both in term of auxiliary contrast function $R$ and of contrast function $\chi$. The results are obtained by considering a number of unknown Fourier coefficients $N_{arm}$ equal to 13x13 and $\beta = 1$ (this latter has been chosen according to Figure 1(e)). Note that all the standard NIE-CSI, H0-CSI (that is the standard CSI) and Y0-CSI (that is the method discussed in [19]) models completely fail in retrieving the B shaped target for this specific example with $\chi = 1.9$. The reconstructions in Figure 2 as well as the reached NMSEs prove that the Y0-NIE-CSI is more convenient than the standard NIE-CSI. Indeed, NIE-CSI is not able to retrieve the target, while the Y0-NIE-CSI corresponds to a reconstruction error of 0.18.

### B. VE − NIE − CSI results

Figure 3 shows the VE-NIE-CSI results both in term of auxiliary contrast function $R$ and of contrast function $\chi$. A number of unknown Fourier coefficients $N_{arm}$ equal to 13x13 and $\beta = 1$ have been considered. In particular, both the results obtained by equipping the VE-CSI cost function (30) with and without the penalty term (26) are shown. As stressed above, the results obtained by means of the standard NIE-CSI are not shown as it completely fails in retrieving the target for this specific example with $\chi$=1.9.

As can be seen from figure 4, the use of NIE model in conjunction of the VE framework allows instead to ensure reliable reconstruction of the target at hand even if no a priori information on the unknown target is exploited in the CSI optimization.

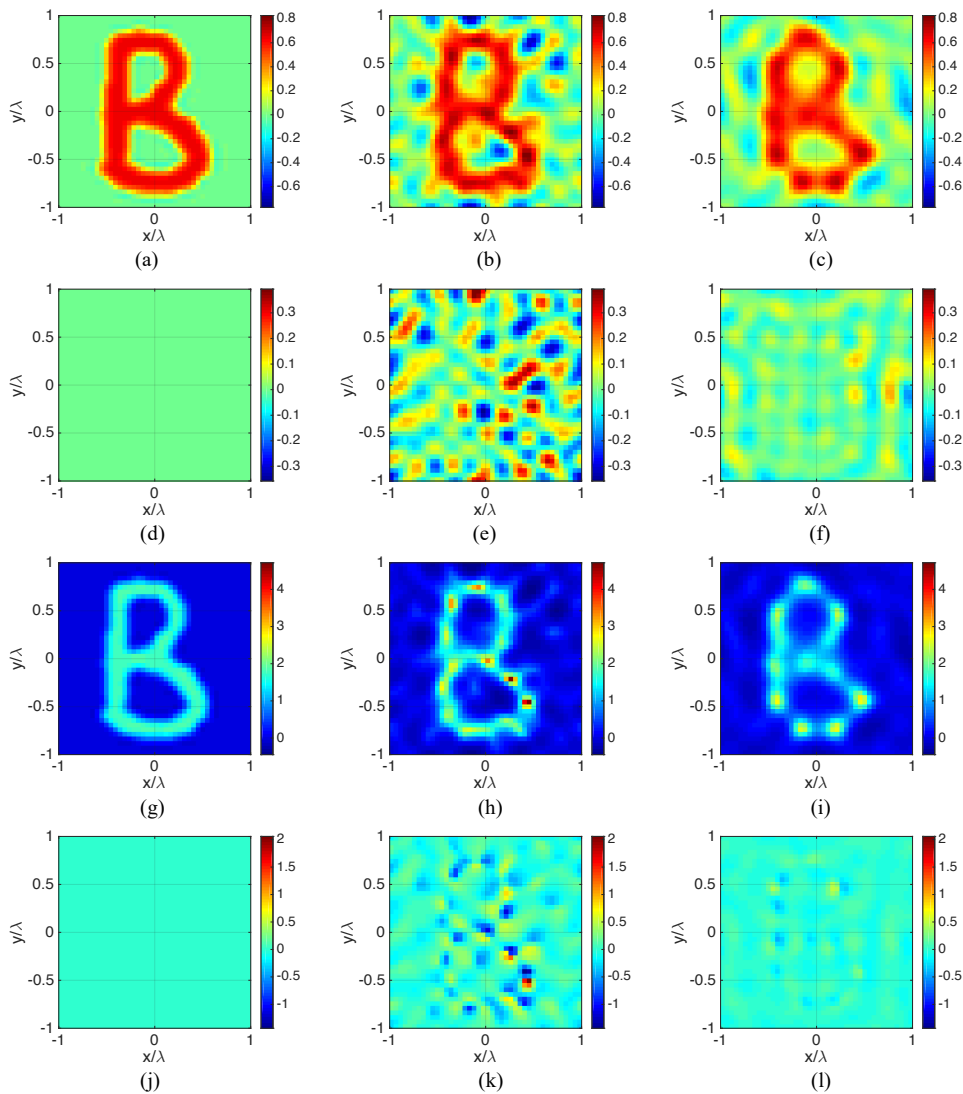

Fig. 3. VE-NIE model. Assessment against numerical data: B target $\chi$=1.9. Real (a) and imaginary (d) parts of the R reference profile. Real (b) and imaginary (e) parts of the retrieved R function (NMSE=0.43). Real (c) and imaginary (f) parts of the retrieved R function (NMSE=0.23) by using the penalty term in (26). Real (g) and imaginary (j) parts of the reference contrast profile. Real (h) and imaginary (k) parts of the retrieved contrast function (NMSE=0.27 - #iter=11687). Real (i) and imaginary (l) parts of the retrieved contrast function (NMSE=0.19 - #iter=1178) by using the penalty term in (26).

## VII. CONCLUSION

In this paper, different strategies, most of them introduced in the last twenty years by our research group, to counteract non-linearity of inverse scattering problem are reviewed and discussed under a unified point of view. First, attention has been devoted to different possible rewritings of the Lippman Schwinger basic equation. Then, we briefly reviewed how a suitable rearrangement of the original data can contribute to condition the scattering phenomenon and simplify the inversion procedure.

For the first part, three different (recent or anyway unusual) rewritings of the scattering equations in homogeneous backgrounds have been reviewed and compared in term of their 'degree of non-linearity' (DNL) [15]. In particular, their DNL, and hence the corresponding difficulties in inversion, are analyzed in term of the function encoding the electromagnetic properties of the targets and of the norm of the corresponding operator entering the state equation. Then, some considerations are given in order to evaluate which model can be more convenient depending on the scenario at hand.

Second, we briefly summarized and discussed (under the same 'rewriting' spirit) the 'Virtual Experiments' (VE) framework [20]-[22]. In particular, we reviewed the possibility to design scattering experiments such to give rise to focused total fields (or contrast sources), or to circular symmetric contrast sources, which can be done according to several different strategies [20]-[22],[40]. Notably, whenever the focusing of the contrast source is actually possible the inverse scattering becomes a kind of 'local' interrogation, so that the effects of nonlinearity are alleviated.

Last, but not least, a couple of possible joint exploitations of the above concepts are introduced, discussed and tested. In particular, the hybrid models and the derived CSI methods, named Y0-NIE-CSI and VE-NIE-CSI, are introduced and tested against a test target which belongs to the MNIST data set [36]. Reconstruction results using very simple regularization techniques, as well as very simple techniques to extract the unknown contrast function from the auxiliary unknowns, confirm the interest and the potentialities of the proposed approaches. Moreover, the results suggest that other hybridizations are also worth to be investigated.

Present work is aimed to confirm the usefulness of the proposed hybridizations when, by virtue of a priori information which is eventually available, more sophisticated regularization techniques can be used.